\documentclass[11pt,oneside,letterpaper]{article}
\usepackage{amssymb}
\usepackage{amsmath}
\usepackage[dvips]{graphicx}
\usepackage{setspace}
\usepackage{amsfonts}
\usepackage{fancyhdr}
\usepackage{xcolor}
\usepackage{graphicx}
\usepackage{rotating}
\usepackage{comment}
\usepackage{color}
\usepackage{cite}
\usepackage{braket}
\usepackage[permil]{overpic}
\definecolor{darkgreen}{rgb}{0,0.5,0}
\definecolor{darkblue}{rgb}{0,0,0.6}
\definecolor{purple}{rgb}{0.4,.2,0.7}
\newcommand{\p}{\partial}

\newcommand{\f}{\frac}

\newcommand{\be}{\begin{equation}}
\newcommand{\ee}{\end{equation}}

\newcommand{\bx}{\bar{x}}

\usepackage[colorlinks=true,citecolor=darkgreen,linkcolor=black,urlcolor=purple]{hyperref}

\usepackage{pdfsync}

\makeatletter
\newcommand*{\defeq}{\mathrel{\rlap{%
                     \raisebox{0.3ex}{$\m@th\cdot$}}%
                     \raisebox{-0.3ex}{$\m@th\cdot$}}%
                     =} 
\makeatother

\def\be{\begin{eqnarray}}
\def\ee{\end{eqnarray}}

\newcommand{\tr}{\textrm{Tr}\,}

\newcommand{\bea}{\begin{eqnarray}}
\newcommand{\eea}{\end{eqnarray}}
\def\ben{\begin{equation}}
\def\een{\end{equation}}

\let\l=\lambda \let\m=\mu   \let\p=\phi \let\r=v
\let\s=\sigma   \let\c=\chi

\let\f=\frac

\def\be{\begin{equation}}
\def\ee{\end{equation}}
\def\ba{\begin{array}}
\def\ea{\end{array}}
\def\del{\partial}

\def\m{{M}}
\def\euv{\epsilon_{\rm uv}}
\def\log{{\rm ln}}

\def\ba#1\ea{\begin{align}#1\end{align}}
\def\bs#1\es{\begin{split}#1\end{split}}

\renewcommand{\p}{\partial}

\allowdisplaybreaks  

\numberwithin{equation}{section}

\def \be {\begin{equation}}
\def \ee {\end{equation}}
	
\def \JM#1 {{\color{blue}  JM: #1 }}
\def \AAl#1 {{\color{red}  AA: #1 }}

\begin{document}
\onehalfspacing

\begin{center}

~
\vskip3cm

{\LARGE \bf { Islands in Asymptotically Flat 2D  Gravity
\\
\ \\
}}

\vskip10mm

Thomas Hartman,$^{1}$\ \ Edgar Shaghoulian,$^{1}$\ \  and Andrew Strominger$^{2}$

\vskip5mm

{\it $^1$ Department of Physics, Cornell University, Ithaca, New York, USA
}  \\
{\it $^2$ Department of Physics, Harvard University, Canbridge, MA, USA } 
\vskip5mm

\vskip5mm

\end{center}

\vspace{4mm}
\begin{abstract}
\noindent
The large-$N$ limit of asymptotically flat two-dimensional dilaton gravity coupled to $N$ free matter 
fields provides a useful toy model for semiclassical black holes and the information paradox. Analyses of the asymptotic information 
flux as given by the entanglement entropy  show that it follows the Hawking curve, indicating that information is destroyed in these models. Recently, motivated by developments in AdS/CFT, a semiclassical  island rule for entropy has been
proposed. We define and compute the island rule entropy for black hole formation and evaporation in the large-$N$  RST model of dilaton gravity and show that, in 
contrast, it follows the unitary Page curve. The relation of these two observations, and interesting properties of the dilaton gravity island rule, are discussed. 
 
 \end{abstract}

\pagebreak
\pagestyle{plain}

\setcounter{tocdepth}{2}
{}
\vfill
\tableofcontents


\section{Introduction}

Many years ago, Gibbons and Hawking \cite{Gibbons:1976ue} showed that the entropy of a black hole can be formally but simply calculated
in the semiclassical limit from the Euclidean path integral. That calculation is now understood, at least in some cases, to correctly enumerate the quantum microstate degeneracy of the black hole. It is a wonderful and deep surprise that semiclassical gravity is smart enough to reproduce this degeneracy. In the intervening decades, the unreasonable efficacy of semiclassical gravity has been repeatedly demonstrated in disparate contexts. Most cases however involve symmetries of one kind or another - such as time translations. Recently  \cite{Penington:2019npb,Almheiri:2019psf,Almheiri:2019hni,Almheiri:2019qdq,Penington:2019kki}
a remarkable proposal has been made invoking semiclassical gravity to analyze the information flow out of black holes. The starting point is the famous and well-understood Ryu-Takayanagi formula \cite{Ryu:2006bv} for anti-deSitter space (AdS),
which semiclassically computes microscopic  quantum entanglement entropies in terms of areas of extremal surfaces. 
Recent proposed generalizations of this formula define an `island rule  entropy' for  time-dependent evaporating black holes in AdS  - and even for those in flat space - involving  a `quantum extremal surface' or QES\cite{Faulkner:2013ana,Hubeny:2007xt,Engelhardt:2014gca}.  The island rule  entropy elegantly reproduces at leading order the Page curve describing unitary information flux, accompanied by a  refined semiclassical picture of the evaporation process.  We hope that this significant progress will provide  important clues to guide us to  a complete microscopic resolution of the information paradox. 

In a separate direction, the process of black hole evaporation has been systematically  studied starting in the early 90s in the context of  the large-$N$ limit of 2D dilaton gravity minimally coupled to $N$ free matter fields. A review can be found in \cite{Strominger:1994tn}. The first such model, known as CGHS \cite{Callan:1992rs},  originated as a dimensional truncation of the near extremal NS fivebrane in string theory, but the CGHS model and its  variants \cite{Bilal:1992kv,Russo:1992ax,Giddings:1992ae,deAlwis:1992hv}  can also be studied as toy 2D black hole models in their own right. These models are of interest because the  large-$N$ limit includes Hawking radiation as well as the backreaction on the metric, and yet is soluble. The quantum state on ${\cal I}^+$, the information flux, or equivalently entanglement entropy as a function of retarded time can be numerically or in some models 
(in particular the RST model \cite{Russo:1992ax}) analytically computed.  The information flux at ${\cal I}^+$ follows the Hawking curve, not the Page curve. The models therefore seem to destroy information.

In assessing this uncomfortable conclusion it is important to realize that the dilaton gravity models, despite some efforts, were never  
derived either as a decoupling limit of string theory or as the large-$N$ limit of an $exactly$ defined model of 2D gravity.  In stringy limits, typically an infinite tower of interacting fields do not decouple.  Since 2D quantum gravity is renormalizable (in fact equivalent to 2D CFT) one might hope to define and study such models in their own right without recourse to a decoupling limit.  The  2D bulk theory is in some cases fully soluble at the quantum level (as a null variant of Liouville theory  \cite{deAlwis:1992hv,Bilal:1992kv}). However, a faithful toy model for black holes requires the dilaton field to be positive, and a boundary is needed to eliminate regions where it becomes negative.  Efforts to implement such a boundary condition and avoid naked singularities  in a self-consistent manner never fully succeeded. 
The apparent destruction of information, together with the absence of any known exact definition at finite $N$, suggest that the large-$N$ dilaton gravity models are not the large $N$ limit of any exact fully consistent 2D quantum  theory of gravity.\footnote{A further reason for doubting the existence of such theories is the exact global flavor symmetry, which is not expected in quantum gravity. It is hard to see how the information about the flavor of the matter which collapses to a black hole  could ever be retrieved from the Hawking radiation at ${\cal I}^+$ .} However the jury is still out concerning the best perspective on these models and their relevance to the information paradox.

It would certainly be of great interest to find $some$ fully consistent 2D toy model to study black hole formation/evaporation.\footnote{Black holes probably cannot be formed  in the $c=1$ matrix model \cite{Karczmarek:2004bw}, and the bulk description  of the  SYK model \cite{Sachdev:1992fk,kitaevfirsttalk} involves a tower of low-mass states. } It remains possible that some modification of CGHS - perhaps involving wormholes, boundary modes or other additional degrees of freedom - exists as an exact quantum theory and obeys the island rule descibed herein. 

Of special interest in this regard are the near extremal NS fivebranes, which contain CGHS plus additional massive fields which survive the scaling limit and are  holographically dual to little string theory \cite{Callan:1992rs,Maldacena:1997cg,Aharony:1998ub}. While these theories are not  fully two-dimensional and do not have $N$ matter fields, they can be semiclassically studied at large black hole mass $\m$. As they do have a holographic dual, it is plausible that the information flux at ${\cal I}^+$  follows the Page curve, and the island rule  entropy as described in this paper agrees with the information flux measured by an asymptotic observer.  It would be very interesting to derive this explicitly.

In this paper we compare in detail in 2D these old and new  semiclassical analyses of black hole evaporation. We  analyze the QES  and associated semiclassical island rule entropy in the  RST model, employing  a definition of the island rule entropy which closely mimics the ones used in AdS  \cite{Almheiri:2019qdq} but lives at ${\cal I}^+$. 
The  explicit expression as a function of retarded time is shown to  reproduce the leading-order Page curve, which   is a stringent consistency check of the island rule. This of course disagrees with the actual measurable  information flux of the RST model which follows the Hawking curve.   We analyze the difference  and try to learn something from it.

More generally, the semiclassical island rule  does not agree with the semiclassical  Hawking calculation for 4D black holes even though they apparently have the same regime of validity. If the former is correct, one would like to find the error in the latter.  Our hope is that  the 2D context considered here proves a useful one for resolving this tension. 

The island rule supplies a surprising canonical map from a time along the QES  curve  in the black hole interior to a retarded time on ${\cal I}^+$.  It implicitly  asserts that when an infalling object crosses the horizon and reaches the QES curve, information about its quantum state becomes available on ${\cal I}^+$.  This assertion reproduces the Hayden-Preskill scrambling time. 
The QES is seen at ${\cal I}^+$ to hit the singularity a time of order one before the evaporation endpoint. The island rule entropy then has no classical gravitational piece, and reduces to the ordinary quantum entanglement of the surface extending from the QES endpoint to 
${\cal I}^+$. This surface magically has zero net quantum entanglement to leading order in $\m$ due to a cancellation of the order $\m$ bulk term with corrections due to the UV cutoff redshift at its boundaries. This 
provides another self-consistency check of the island rule.

This paper is organized as follows. Section 2 is a lightning review of the needed features of 2D asymptotically flat dilaton gravity and the RST model.  The linear dilaton, eternal black hole and evaporating black hole solutions are presented, followed by  the microscopic entanglement entropy and the information flux at
${\cal I}^+$ for the evaporating black hole. Section 3 begins with the definition of the island rule entropy, which  adds to the asymptotic entanglement entropy an island region which ends at the QES with an area-law type gravitational contribution. This formula is then applied to the evaporating black hole and shown to have several remarkable properties, including reproducing the Page curve and the scrambling time.  Section 4 considers the 
eternal black hole, which has a mathematically simplified, if less physically sharp, version of the information paradox: the entanglement entropy of Hawking quanta outside the black hole grows without bound and eventually exceeds the Bekenstein-Hawking black hole entropy. This paradox is beautifully resolved by assuming the island rule.  In the appendix we consider the gravitational replica calculation and show that the local equations of motion fix the location of the QES, although we will not construct the wormhole saddles globally.

 Other recent work addressing islands and information recovery in black holes includes \cite{Zhao:2019nxk, Akers:2019nfi, Rozali:2019day, Chen:2019uhq, Bousso:2019ykv, Almheiri:2019psy, Chen:2019iro, Laddha:2020kvp, Mousatov:2020ics, Kim:2020cds, Saraswat:2020zzf, Chen:2020wiq, Marolf:2020xie, Verlinde:2020upt, Jana:2020vyx, Giddings:2020yes, Liu:2020gnp, Pollack:2020gfa, Balasubramanian:2020hfs}. As this work was nearing completion two  papers appeared with overlapping results \cite{Gautason:2020tmk,Anegawa:2020ezn}.

\section{Review of the RST model}

This section contains a lightning review of relevant features of the RST model \cite{Russo:1992ax}, largely following the conventions of \cite{Fiola:1994ir}
to which we refer the reader for further details. 

\subsection{Large-$N$ action and equation of motion}
The classical action of the RST model  is\footnote{More generally, replacing $S_{\text{CFT}}$ with any family of 2D CFTs with central charge $c=N$ minimally coupled to gravity would lead to few changes in the following discussion. The $\phi R$ term in \eqref{rstaction} distinguishes the RST from the CGHS model \cite{Callan:1992rs} and was added to restore the conserved current $\partial_\mu (\rho-\phi)$ broken by the anomaly. The existence of this conserved current conveniently simplifies many formulae but does not affect the basic structure, see e.g. \cite{Piran:1993tq}. Indeed for our primary focus of the large mass limit there is no difference. }
\bea\label{rstaction}
S_{cl} &=& \f{1}{2\pi}\int d^2 x \sqrt{-g} \left[e^{-2\phi}(R+4(\nabla \phi)^2+4\l^2) 
-\f{N}{24}  \phi R\right]+S_{\text{CFT}}\cr S_{\text{CFT}}&=& -\sum_{k=1}^N\f{1}{4\pi}\int d^2 x \sqrt{-g}(\nabla f_k)^2   .
\eea
To avoid equation clutter we henceforth choose units  in which 
\be \l=1.\ee
In the quantum effective action, the conformal anomaly contributes a term  \be S_{\rm anom} = -\frac{N}{96\pi} \int d^2 x \sqrt{-g}R \Box^{-1}R.\ee   The large-$N$ limit is  $N\to \infty$ with $Ne^{2\phi}$ held fixed. It described  by the sum \be S_{\rm eff}=S_{\rm cl}+S_{\rm anom}+S_{\text{CFT}}.\ee
This action simplifies in conformal gauge $ds^2 = -e^{2\rho}dx^+ dx^-$, with $x^{\pm} = x^0\pm x^1$ and $R_{+-}=-2\del_+\del_-\rho$. In these coordinates the large-$N$ effective action  becomes
\bea\label{rsteff}
S_{\rm eff}&=&\f{1}{2\pi}\int d^2 x\, \left[e^{-2\phi}\left(4\p_+\p_- \rho-8 \p_+\phi \p_-\phi+2e^{2\rho}\right) +\f{N}{6}(\rho-\phi) \partial_+\partial_-\rho \right] +S_{\text{CFT}}, \cr
S_{\text{CFT}}&=& \f{1}{2\pi}\sum_{k=1}^N\int d^2 x \del_+ f_k\del_- f_k  
\eea
It is convenient to define 
\be\label{omegachi}
\Omega = {12 \over N}e^{-2\phi}+\f{\phi}{2}-\f 1 4 \log {48 \over N}\,,\qquad \chi = {12 \over N}e^{-2\phi}+\rho-\f{\phi}{2}+\f 1 4 \log {3 \over N}.
\ee
The resulting action 
\be\label{bilca}
S_{\rm eff}=\f{N}{12\pi}\int d^2 x \,\left(\partial_+ \Omega\partial_- \Omega -\partial_+\chi \partial_-\chi + e^{2\chi-2\Omega}\right)+S_{\text{CFT}}
\ee
then scales with an overall factor of $N$ in our large-$N$ limit where $\Omega$ and $\chi$ are held fixed. 

Note that for real $\phi$, $\Omega$ obeys \be\label{dsc} \Omega \ge {1 \over 4}.\ee In order to identify $S_{\rm eff}$ as a theory of gravity with black holes, it is essential that the restriction of real $\phi$ be enforced so that the coefficient of the Einstein action is positive.  When 2D dilaton gravity is constructed as  a spherical dimensional reduction of higher-dimensional gravity, $\Omega$ is  the (quantum) area of the spheres, and the condition \eqref{dsc} marks the origin of the spacetime where the spheres shrink to zero.  If we don't make such a restriction, \eqref{bilca} 
describes  a soluble, null variant of Liouville theory   \cite{deAlwis:1992hv,Bilal:1992kv}, but it it has no singularities and no black holes.\footnote{Some versions  of JT gravity have this feature and hence may need supplemental restrictions to provide good models for black hole physics.}  Furthermore, without this restriction the collapsing matter configurations described below will radiate forever to negative infinite mass, whereas with the restriction we can motivate a pasting of the linear dilaton vacuum at the evaporation endpoint, i.e. the point where the $\Omega = 1/4$ curve turns from spacelike to timelike. This cuts off the evaporation. In order  to have a theory with black holes,   we therefore impose \eqref{dsc} as in \cite{Russo:1992ax,Fiola:1994ir}. This condition looks sensible at large $N$, but certainly spoils the full quantum solubility.\footnote{ Whether any such constraint makes sense at finite $N$ is an open question, studied in \cite{Strominger:1994xi,Polchinski:1994zs}.}  We further impose reflecting boundary conditions on the $N$ quantum matter fields $f_k$  which descends from their reflection  through the origin in higher dimensions.

The $S_{\rm eff}$ equations of motion for $\chi$ and $\Omega$ are $\partial_+ \partial_- \Omega = \partial_+ \partial_- \c= - e^{2 \c - 2 \Omega}$, so $\chi = \Omega + \alpha^+(x^+) + \alpha^-(x^-)$. The residual on-shell conformal diffeomorphism symmetry of conformal gauge allows us to choose coordinates in which  \be\label{cog} \chi = \Omega.\ee 
For reasons which will become apparent we refer to these as Kruskal coordinates and will denote the non-scalar quantities $\chi$ and $\rho$ in Kruskal gauge as $\chi_K$ and $\rho_K$. \eqref{cog} implies \bea \label{rp} \phi&=&\rho_K - \frac{1}{2}\log \frac{N}{12}\cr \Omega &=& e^{-2\rho_K} + \frac{1}{2}(\rho_K  - \log 2).\eea The equation of motion in this gauge is simply 
\be\label{eom}
\partial_+\partial_- \Omega = -1\,.
\ee
There are also constraint equations, which are given by varying \eqref{rstaction} with respect to $g_{\pm \pm}$ before fixing to conformal gauge:
\be\label{constraint}
\partial_\pm^2 \Omega = -T^{f}_{\pm \pm}-t_\pm \\,
\ee
where 
\be\label{rds}T^f_{\pm\pm}= \f{6}{N}\sum_{k=1}^N\int d^2 x \del_\pm f_k\del_\pm f_k \ee
is the matter stress tensor rescaled by a factor of $12 \pi \over N$ for notational convenience.
 $t_\pm$ depends on the  choice of coordinates prescribed  for normal ordering. If we choose ${x'}^+(x^+)$instead of $x^+$, $t_+$ shifts by the Schwarzian (see e.g. \cite{Fiola:1994ir})
\be\label{sch} t'_+=({\del'}_+ x^+)^2t_++\sqrt{{\del'}_+x}{\del'}_+^2\sqrt{\del_+{x'}^+},\ee
with a similar relation for $t_-$.   In the black hole collapse studied below the ambiguity will be uniquely fixed by 
demanding the absence of incoming particles in inertial frames in the far past.

\subsection{Solutions}
The general solution to \eqref{eom}-\eqref{constraint} can be obtained by integration, although we will only need the following special cases. 
\subsubsection{Linear dilaton vacuum}
\begin{figure}
\begin{center}
\includegraphics[scale=1.0]{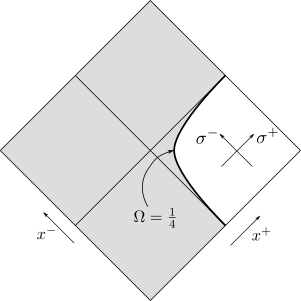}
\end{center}
\caption{\small Linear dilaton vacuum. The physical spacetime is the unshaded region, with a boundary at $\Omega = \frac{1}{4}$.\label{fig:vacuum}}
\end{figure}

The vacuum of the theory, known as the linear dilaton vacuum, in  Kruskal coordinates \eqref{cog} is 
given by 
\be\label{ldv}
\Omega = -  x^+ x^- - \f 1 4 \log(-4  x^+ x^-)\,
\ee
with the classical matter stress tensor $T^f_{\pm\pm}=0$. 
The physical region of the spacetime with $\Omega\ge{1 \over 4}$ is covered by the coordinate region with $x^+x^-\le {1 \over 4}$
and $\pm x^\pm>0$, as illustrated in fig. ~\ref{fig:vacuum}. Using \eqref{cog}, this implies the 2D Minkowski line element 
\be ds^2={dx^+dx^-\over x^+x^-}.\ee  From the fact that $\p_\pm^2\Omega={1 \over 4(x^\pm)^2}$, we conclude that \be t_\pm=-{1 \over 4(x^\pm)^2}.\ee
We will also employ
 ``Minkowski coordinates"  defined by \be x^\pm = \pm e^{\pm \s^\pm}\ee 
 in which the line element takes the standard flat form
 \be
ds^2 = -d\s^+ d\s^-,
\ee 
the dilaton is linear 
\be \phi=\f 1 2 \left(\s^--\s^+-\log \f{N}{12}\right),\ee
and $\Omega=\chi+\f 1 2(\s^--\s^+)$.
The standard Minkowski vacuum is annihilated by Minkowski modes which are negative frequency with respect to Minkowski time $\s^++\s^-$. Using the transformation law \eqref{sch} we find that $t_\pm=0$ in the Minkowski frame and \eqref{ldv} hence corresponds to the quantum fields $f_k$ in the standard Minkowski vacuum.

 \subsubsection{Eternal black hole}

The solution for the eternal black hole  (see fig. ~\ref{fig:eternalBH}) is
\be\label{eternalomega}
\Omega = - x^+ x^- + \m ,
\ee 
with  $T^{f}_{\pm \pm} =t_\pm =0$. To make sure the singularity at $\Omega = 1/4$ is spacelike we choose $\m > 1/4$. The event horizon of the black hole is at $x^+ x^- = 0$. The conformal factor is obtained from \eqref{omegachi} in Kruskal coordinates as $\rho_K = 2\Omega + \log 2 + \f 1 2 W_{-1}(-e^{-4\Omega})$, where $W_{-1}$ is the product logarithm function. In Minkowski coordinates, the metric near infinity is $ds^2 = -d\s^+ d\s^-$. Normal ordering in these coordinates leads to an energy observed by asympotically inertial observers,
\be\label{tyl} t_+(\s^+)={1 \over 4}.\ee
This arises from outgoing thermal radiation flux at temperature 
\be T={1 \over 2 \pi}.\ee
Using the first law ${\del S_{BH}\over \del E}={1 \over T}$ and $E={N\m \over 12\pi}$
one finds the black hole entropy
\be S_{BH}={N\m \over 6}.\ee

\subsection{Evaporating black hole in RST}
A beautiful feature of the RST model is that the  formation and evaporation of a black hole can be described exactly and analytically in the semiclassical limit, at least up to the endpoint of the evaporation process. The corresponding solution is
\be\label{omegaevap}
\Omega = - x^+ x^- - \frac{1}{4}\log(-4 x^+ x^-) - \m(x^+ - 1) \Theta(x^+ -1 ) .
\ee
The spacetime is illustrated in fig. ~\ref{fig:evapX} in Kruskal coordinates \eqref{cog}, and the conformal diagram appears below in fig. ~\ref{fig:evap-big}.
\begin{figure}
\begin{center}
\includegraphics[scale=1.0]{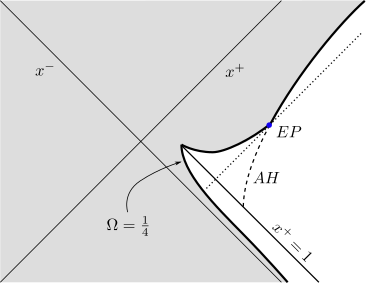}
\end{center}
\caption{\small Evaporating black hole in Kruskal coordinates. The apparent horizon $AH$ is on the dashed line and the evaporation endpoint is marked $EP$. The dotted line prior to $EP$ is the event horizon. \label{fig:evapX}}
\end{figure}
There is an incoming matter shockwave taken to be at $x^+ = 1$ (or $\sigma^+=0$ in Minkowski coordinates) that creates the black hole. Using the constraint equation \eqref{constraint}, the matter stress tensor is 
\be T^f_{++}=\m \delta(x^+ - 1), ~~~T^f_{--}=0, ~~~t_\pm = -\frac{1}{4(x^\pm)^2}.\ee
The curve $\Omega = \frac{1}{4}$ is the edge of the spacetime. This curve is initially timelike, as in the linear dilaton vacuum; here we impose reflecting boundary conditions on the matter fields. When the shockwave hits the boundary, the curve $\Omega = \frac{1}{4}$ turns spacelike, and is interpreted as the black hole singularity.  After the black hole completely evaporates at the endpoint it becomes a timelike boundary again.

The solution has an apparent horizon defined by $\p_+ \Omega = 0$,\footnote{Interpreting $\Omega$ as the generalized quantum area, this is the boundary of the region of trapped spheres.} which is the outer boundary of the evaporating black hole. This is the curve
\be
x^+ (  x^-  +{\m}{} )= -\frac{1}{4 } \ 
\ee
for $x^+>1$.
The evaporation ends where the apparent horizon meets the singularity. This occurs at the endpoint denoted $P_{EP}$ with
\be\label{ep}
x_{EP}^- = \frac{\m }{e^{-4\m } - 1} , \quad
x_{EP}^+ = \frac{1}{4\m}(e^{4 \m }-1) .
\ee
This value of $x^-$ also defines the event horizon. 

After the endpoint $P_{EP}$, the singularity emerges from behind the horizon, the radiation continues 
and, problematically, the mass of the spacetime becomes arbitrarily negative. RST propose a new, modified  
``thunderbolt" boundary condition at $P_{EP}$ which returns the system to the linear dilaton vacuum, matching along $x^-_{EP}$.
However, the details of this do  not concern us here because all of our considerations address the behavior of the theory $prior$ to the future light cone of $P_{EP}$. In particular we wish to understand how much information is returned before the retarded time  $x^-_{EP}$.

Near ${\cal I}^-$, the geometry is in the linear dilaton vacuum, so the metric is $ds^2=-d\sigma^+ d\sigma^-$ in the incoming Minkowski coordinates defined by 
\be\label{imc}
 x^{\pm} = \pm e^{\pm  \sigma^\pm}  .
\ee
The (rescaled) initial mass of the black hole is equal to the energy of the incoming shockwave. In Minkowski coordinates
\be
\m = \int d\sigma^+ T_{++} ={12 \pi \over N}M_{ADM},
\ee
where $M_{ADM}$ is the canonical ADM mass. 
The initial black hole entropy is
\be\label{sinit}
S_{BH} = \frac{N \m }{6} =2\pi M_{ADM}.
\ee
Near ${\cal I}^+$, the conformal factor in Kruskal coordinates is 
\be\label{confplus}
e^{2\rho_K} = - \frac{1}{x^+ (x^- + \m)} + {\cal O}\left( \frac{1}{(x^+)^2}\right) .
\ee
In the outgoing Minkowski coordinates $\tilde \s^\pm$ defined in terms of Kruskal coordinates by 
\be \label{omc} x^+=e^{\tilde\s^+},~~~x^-+\m=-e^{-\tilde \s^-},\ee
the metric near ${\cal I}^+$ takes the manifestly  flat form 
\be ds^2=-d\tilde{\sigma}^+ d \tilde{\sigma}^-\left(1+{\cal O}(e^{-\s^+})\right).\ee
Comparing to \eqref{ep} we note that the endpoint occurs at the (affine) retarded time 
\be \label{rep} \tilde \s^-_{EP}=\log {e^{4\m}-1\over \m} \sim 4\m-\log \m+\cdots ,\ee
where the corrections are suppressed by powers of $e^{-4\m}$.
To leading order the evaporation time of $4\m$ matches the time it takes to radiate away a total mass $\m$ with an $\m$-independent energy flux 
$t_+={1 \over 4}$ in \eqref{tyl}. 

\subsubsection{Entanglement entropy}

\begin{figure}
\begin{center}
\includegraphics[scale=1.0]{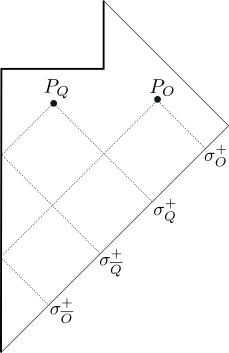}
\end{center}
\caption{\small Points are labeled by $(\sigma^+_P, \sigma^+_{\overline{P}})$, where $\sigma^+_{\overline{P}}$ is the value of $\sigma^+$ for the  image point obtained by reflecting across the timelike boundary.\label{fig:reflect}}
\end{figure}

The large $N$ limit includes information about the quantum state of the $N$ free scalars on ${\cal I}^+$. This is a pure state on any everywhere spacelike or null slice which 
stretches from spatial infinity to the timelike portion of the singularity -- the `origin' -- where $\Omega={1 \over 4}$ and reflecting  boundary conditions are imposed on the $f_k$. Such slices are all prior to the future light cone of the endpoint. 
The quantum state on a subregion $A$ of such a slice is a density matrix $\rho_A$ obtained by tracing over the quantum  state on the slice outside of $A$.  The entanglement entropy \be S_{\rm ent}(A)=-\mbox{tr}\, \rho_A\log \rho_A \ee
can be computed exactly \cite{Fiola:1994ir} when one end of $A$ is at spatial infinity, the other at an interior point $P_O=( \s^+_O,\s^-_O)$ and the quantum fields are in their vacuum state on ${\cal I}^-$.  $P_O$ will often be taken to be the location of an Observer. The result is \cite{Fiola:1994ir}
\be \label{snt} S_{\rm ent}(A)={N \over 6}\big[ \rho_{\rm in}(\s^+_O,\s^-_O)+\log {\s^+_O-\s^+_{\bar O} \over \euv}\big].\ee
Here  $\rho_{\rm in} $ is the conformal factor evaluated in the incoming Minkowski coordinates $\sigma^\pm$ used to define the vacuum, and $\s^+_{\bar O} $ is the incoming coordinate of the null line which reflects off the boundary and arrives at $P_O$ (see fig. \ref{fig:reflect}), which obeys 
\be \s^+_{\bar O} =\s^-_O-\log 4.\ee  $\euv$ is a UV regulator required by the divergent entanglement of short-wavelength modes across the point $P$ on the spacelike slice. The factor of $\rho_{\rm in}$ appears because one wants the UV regulator to have the same fixed proper length at any slice endpoint in the spacetime. 

\eqref{snt} in fact applies to any CFT with the simple replacement of $N$ by the central charge $c$.  (Of course to take a large-$c$ limit we need a sequence of theories labeled by arbitrarily large $c$.) A more involved formula for more general choices of the region $A$ will be introduced below when it is needed in the definition of the island rule entropy. 

If $P_O$ is chosen to be a point on the apparent horizon, \eqref{snt} is the entropy of the quantum fields outside the black hole. If $P_O$ is chosen to be a point on ${\cal I}^+$, 
it is the entanglement entropy of the portion of the outgoing state prior to $P_O$ with that after $P_O$. This entanglement entropy is what is physically measured by an inertial asymptotic detector that cannot detect modes shorter than $\euv$. In terms of the outgoing retarded Minkowski coordinate $\tilde \s^-_O$ (eq. \eqref{omc}) of $P_O$, equation \eqref{snt} becomes 
\be\label{sep}S_{\rm ent}(\tilde \s^-_O)=  {N \over 12}\log(1+\m e^{\tilde \s^-_O})+{N \over 6}\log{\s^+_O-\s^+_{\bar O} \over \euv}.\ee
Since $\s^+_O\to \infty$ on ${\cal I}^+$, the second term contains both IR and UV  divergences. These are independent of the retarded time of the observer and so are not
associated to outgoing Hawking  flux.  
At late times the entropy grows as ${N\over 12}\tilde \s^-_O$, precisely as expected for a thermal flux of $N$
scalars at temperature $T={1 \over 2\pi}$. At the evaporation endpoint the net entropy in the Hawking radiation, which is the finite change in the entanglement entropy,  is
\be \label{stot} \Delta S_{\rm ent}(\tilde \s^-_{EP})=  {N M\over 3}.\ee
This is exactly twice the initial entropy \eqref{sinit}, indicating the evaporation process is not adiabatic. 

For the usual reasons, it is hard to reconcile \eqref{stot} with quantum mechanical unitarity. It appears that, to the extent that it is well-defined, the RST model (and its cousins) destroys information.\footnote{In general there is a concern (or hope) that off-diagonal terms in the Hawking radiation density matrix  could conspire to correct the leading semiclassical (or large-$N$) result. This is clearly impossible in the RST model -- which is by definition on a fixed topology -- due to the flavor symmetry of the $N$ bosons. They are strictly decoupled and therefore the flavor information cannot be encoded in the off-diagonal elements of the density matrix. } Hence we seek some kind of modified semiclassical theory with a different entropy flux. This brings us to the next section. 

\section{The island rule for an evaporating black hole }

\begin{figure}
\begin{center}
\includegraphics[scale=0.8]{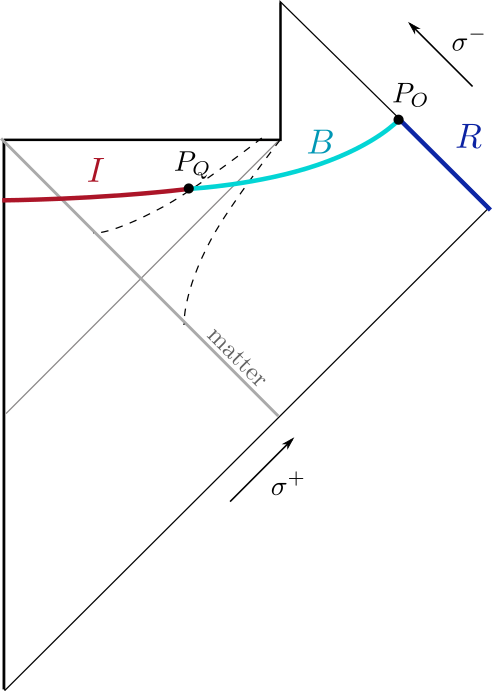}
\end{center}
\caption{\small Island in the evaporating black hole. The Observer at $P_O$ collects the radiation in region $R$. The island is $I$, and $I \cup B \cup R$ is a Cauchy slice. The endpoint of the island is the quantum extremal surface $P_Q$. The dashed lines are the apparent horizon and the curve of extremal surfaces. \label{fig:evap-big}}
\end{figure}

\subsection{Quantum extremal surface}\label{ss:qesentropy}
The island rule  associates an entropy to a point $P_O$ which  involves an extremization of a function over the location $P_Q$ of the quantum extremal surface. The formula for the island rule entropy is \cite{Penington:2019npb,Almheiri:2019psf,Almheiri:2019hni} 
\be\label{islandrule3}
S_I(P_O) =   \min \mbox{ext}_{P_Q}\left[ S_{\rm gen}(I \cup R) \right] \ .
\ee
$P_Q$ can consist of several points in general, but in this section $I$ is an ``island" region which extends from the timelike origin to a single point $P_Q$, and $R$ is the region from spacelike infinity to $P_O$ through which Hawking radiation passes (see fig.~\ref{fig:evap-big}). We also use $B$ to denote the region between $P_Q$ and $P_O$. We restrict the points so that  $I\cup B\cup R$ is a complete everywhere spacelike or null Cauchy surface.  
The generalized entropy is 
\be\label{sgen}
S_{\rm gen}(I \cup R) = S_{\rm grav}(P_Q)  + S_{\rm ent }(I \cup R)    \ .
\ee
The first term $S_{\rm grav}$ is the Bekenstein-Hawking entropy. In the RST model this is \cite{Fiola:1994ir}
\begin{align}\label{sgrav}
S_{\rm grav}  &= \frac{N}{6}\left( e^{-2\rho_K} - \frac{\rho_K}{2}\right) + \frac{N}{24}(\log 4 - 1) + \frac{N}{6} \log \euv\ ,
\end{align}
where $\rho_K$ is in Kruskal coordinates.\footnote{We note $ \rho_{\rm in}= \rho_K  + \frac{1}{2}(\sigma^+ - \sigma^-) $.} The $\log \euv $ term in \eqref{sgrav} will cancel against the UV divergence of the matter entropy coming from the point $P_Q$ when computing $S_{\text{gen}}$.  The point $P_Q$ which extremizes the right hand side of \eqref{islandrule3} is the quantum extremal surface.  If there are multiple extrema the generalized entropy is computed by picking the solution which minimizes the generalized entropy. The $S_I$ defined in \eqref{islandrule3} is a natural proposed generalization to the asymptotically flat 2D context of formulae which can be derived in some AdS contexts,  but it has not been identified as an asymptotic observable quantity  in the standard formulation of the RST model. One may seek to modify the RST model so that \eqref{islandrule3} is a physical observable. This would be a major achievement.

The matter term $S_{\rm ent }({I \cup R})=S_{\rm ent }(B)$ is given by a formula which is more complicated than \eqref{snt} because $B$ does not terminate at spatial infinity. Assuming the scalars are compactified at the self-dual radius, one finds \cite{Calabrese:2009ez}
\be\label{sdboson}
S_{\rm ent}(I \cup R)  = \frac{N}{6}\log\left[ 
\frac{
d^2(P_O,P_Q) d(P_O, P_{{\bar{O}}}) d(P_Q, P_{\bar{Q}}) }{
 \euv^2 
 d^2(P_Q, P_{\bar{O}})
 e^{-\rho_{\rm in}(P_Q) - \rho_{\rm in}(P_O)}
 }
 \right]
\ee
with $d^2(P,P') \equiv (\sigma_P^+ - \sigma_{P'}^+)(\sigma_{P'}^- - \sigma_{P}^-)$ the Minkowski distance. When the conformal cross ratio of the four endpoints satisfies $\frac{d(P_Q, P_O)d(P_{\bar{O}}, P_{\bar{Q}})}{d(P_Q, P_{\bar{O}}) d(P_{\bar{Q}}, P_O)} \ll 1$, this can be approximated by
\be\label{holoS}
S_{\rm ent}(I\cup R) = \frac{N}{6}\log  \frac{ - (\s^+_{\bar Q}  - \s^+_{\bar O} )(\s^+_{Q}  - \s^+_{O} ) }{\euv ^2 e^{-\rho_{\rm in}(P_Q) - \rho_{\rm in}(P_O) } } \ .
\ee
In fact, this latter equation applies to any compact CFT by an OPE argument \cite{Calabrese:2009qy}. We are assuming the scalar is compact (with $O(1)$ radius) to avoid subtleties with the zero mode that would modify this formula \cite{Fiola:1994ir,Calabrese:2009ez}.

This is the CFT entropy in the vacuum state. To describe a  shockwave we may use  a coherent state of the matter fields which does not affect the entropy \cite{Fiola:1994ir}, so this formula also applies to the evaporating black hole.

\subsection{Evaporating black hole}
In this subsection we compute the island rule entropy for the case of an evaporating black hole. We will focus on the regime where the entanglement entropy is given by \eqref{holoS}, since it can be checked that the this is always valid for the island.
Adding \eqref{sgrav} to \eqref{holoS} and sending  $P_O \to {\cal I}^+$ one finds  
\begin{align}\label{he}
S_{\rm gen}(I \cup R) &= \frac{N}{6}\left[\Omega(P_Q) -{1 \over 4}+  \frac{1}{2}(\s^+_{Q}  - \s^+_{\bar Q} ) + \log (\s^+_{\bar Q}  - \s^+_{\bar O} ) 
\right]  \\
& \qquad \qquad \qquad  + \frac{N}{6}\log \left( \frac{\s^+_{O} }{\epsilon_{\text{uv}}\sqrt{1-4\m e^{ \s^+_{\bar O} } }} \right). \notag
\end{align}
Note that the UV and IR divergences $\log \f{\s^+_O}{\epsilon_{\text{uv}}}$ from the point $P_O$ remain. However they do not affect the extremization. 
In fact only the first line depends on the point $P_Q$. The extremality conditions $\p_{\s^+_{\bar Q} }S_{\rm gen} = \p_{\s^+_{Q} } S_{\rm gen} = 0$ are
\be
\m(\s^+_{\bar Q}  - \s^+_{\bar O} ) = e^{- \s^+_{Q} }  = {4\m} - e^{- \s^+_{\bar Q} }  \ .
\ee
In Kruskal coordinates,
\be\label{kruskalQES}
x_Q^+  (x_Q^- + {\m}) = \frac{1}{4} \ , \ee
\be \label{xqo}x^-_O=x^-_Qe^{4(x^-_Q+\m )\over \m}.
\ee
A solution to these equations defines a quantum extremal surface (QES). In this case there are two solutions  depending on the chosen branch cut:
\begin{align}
x_Q^ - = \f{\m}{4}W_{-1}\left(\f{4x_O^- }{e^4\m}\right)\,,\qquad x_Q^+ = \f{1}{4(\m+x_Q^-)}\,;\label{domQES}\\
x_Q^ - = \f{\m}{4}W_0\left(\f{4x_O^- }{e^4\m}\right)\,,\qquad x_Q^+ = \f{1}{4(\m+x_Q^-)}\,.\label{subdomQES}
\end{align} 
$W_k(x)$ is the product logarithm function with branch cut indicated by $k$.

Once we have found the QES, and solved for $\s^\pm_Q$, $S_I (P_O)$ in  \eqref{islandrule3} becomes a function only of the retarded observer time $\tilde \s^-_O$. The explicit expression however involves the product logarithm function and is unilluminating. A simple formula for large mass is given below.

Recall that the apparent horizon is on the curve $x^+  (x^- + {\m}{}) = - \frac{1}{4}$, and the event horizon is slightly before $x^-+ {\m}{}=0$ . Therefore the QES sits inside the event horizon, and lies on a curve obtained by reflecting the apparent horizon across the event horizon.

Because all time foliations are on equal footing in general relativity, one generically expects that there is no canonical way to associate a boundary or retarded time with a point in the interior  --  especially if the interior is inside the black hole.  Contrary to this expectation, \eqref{xqo} explicitly associates a retarded time to the QES inside the black hole. Such an association might  seem required in any mechanism which pulls information out of the black hole.

\subsubsection{When does the QES hit the singularity?}

Eventually the QES always collides with  the singularity. After this point, we do not know how to define the island rule, and the calculation breaks down. The collision  occurs where the QES curve given by the first equation in \eqref{kruskalQES} hits the singularity $\Omega = \frac{1}{4}$, namely
\be
x^-_{QE} = - \frac{\m }{1 + e^{2-4\m }} \ ,
\ee
which is greater than $x^-_{EP}$.
Using the  map \eqref{xqo} between QES time and retarded time and taking $\m$ to be large, the collision `appears' to an observer on ${\cal I}^+$
at retarded time 
\be
\tilde\s_{OQE}^- \approx 
4\m-\log \m -\log 3 -2 +\cdots
\ee 
where the corrections are suppressed by powers of $e^{-4\m}$. Comparing to expression \eqref{rep} we see that the collison appears (for large mass) before 
 the endpoint by an amount
 \be \Delta \tilde \s^-=-2-\log 3. \ee
 The conclusion is that even for large mass, the breakdown occurs only a time of order one before the endpoint. We do not know how to extend the analysis beyond this time, but this will not prevent us from reproducing the Page curve to leading order at large $\m$.

 One may check directly at order $\m$ that the island rule entropy vanishes when the QES hits the singularity.  Interestingly, there is no order $\m$ gravitational contribution ($S_{\rm grav}$) to the island rule entropy, because the end of the island is on the singularity where the area is small.  Therefore the entire expression can be interpreted as CFT entanglement entropy of $I\cup R$ or equivalently $B$. The entanglement of the projection of region $B$ back to ${\cal I}^-$ is represented by the $\log (\s^+_{\bar Q}  - \s^+_{\bar O})$ term in \eqref{he}. This projected region is a narrow sliver of affine length $e^{-4M}$, due to the tremendous blueshift. Therefore this term gives a $negative$ contribution of $-{2N\m \over 3}$ to the entanglement entropy. This is possible because we have subtracted the cutoff dependence, which is operationally equivalent to taking $\euv=1=\lambda$. The (subtracted) entanglement of a region smaller than the cutoff then becomes negative. However this negative contribution is exactly cancelled by the $N\rho_{\text{in}}(\s) \over 6 $ factors at the two ends of the interval $B$ (the third and last factors in \eqref{he}), each of which give a factor of $N\m \over 3$.

Of course the combined region $I\cup B\cup R$ also has zero entanglement entropy since it is a Cauchy slice. It is curious that removing the region $B$ does not change the entanglement entropy. This is a very special property of the exact point $\s^-_{OQE}$ where the QES is seen to hit the singularity. Had we waited a time of order one the $N\rho_{\text{in}}(\s) \over 6 $ factors would not have changed much, but the projected interval on ${\cal I}^-$ would have become exponentially smaller, resulting in a negative entanglement entropy of order $M$.  This provides a stringent consistency  test of the island rule.

\subsection{Page curve}
In this subsection we derive the Page curve to leading order at large $\m$, dropping subleading terms of order $\log M$. 

According to the island rule \eqref{islandrule3}, the entropy as a function of $\tilde \s^-_O$ is obtained by first finding all surfaces obeying the quantum extremality condition, then choosing the one with minimal entropy. One choice is the trivial surface, with no island. The island rule entropy in this case simply equals the quantum entanglement entropy $S_{\rm ent}(\tilde \s^-_O   )$ in equation \eqref{sep}, which follows the Hawking curve.  For a single island, the only other choice is \eqref{domQES}. Therefore the island rule entropy is 
\be
S_I(\tilde \s^-_O) = \min\left[ S_{\rm ent}(\tilde \s^-_O   ) \ , \ S_{\rm gen}(I \cup R) \right]
\ee
where the second term is evaluated at the  the extremum given by \eqref{domQES}.

For $M$ large and $\tilde \sigma_O^-$ of order $M$ we can solve to leading order for the location of the QES using  \eqref{kruskalQES} and \eqref{xqo}. One finds 
\be\label{approxQ}
x^+_Q \approx \frac{3}{4} e^{ \tilde{\sigma}_O^- } , \qquad
x^-_Q +\m\approx \frac{1}{3}e^{- \tilde{\sigma}_O^-} \ .
\ee
Plugging into the island rule one finds
\be
S_I = \min\frac{N}{24}\left[ 2\tilde{\sigma}_O^- \ , \
\tilde{\sigma}^-_{EP} - \tilde{\sigma}_O^- \right]+{N \over 6}\log \f{\s^+_O}{\epsilon_{\text{uv}}}
\ee
where $\tilde{\sigma}^-_{EP}  \approx 4 \m $ is the evaporation endpoint. 

\begin{figure}
\begin{center}
\begin{overpic}[]{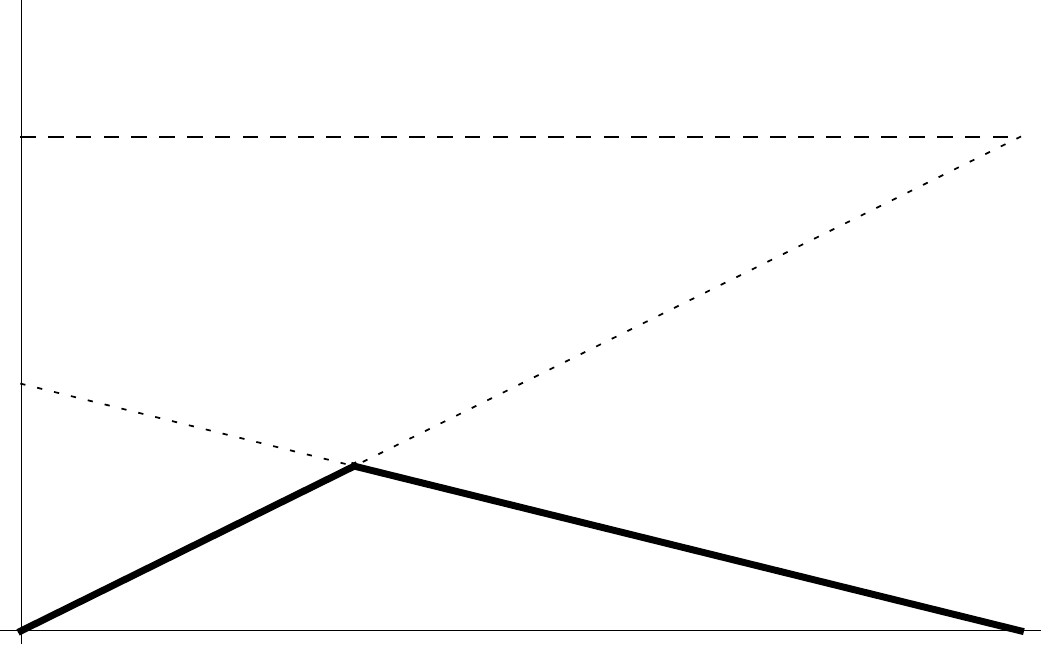}
\put (-45, 580) {$\Delta S$} 
\put (100,500) {$2S_{\rm BH}$}
\put (400,-30) {affine time $\tilde{\sigma}_O^-$}
\put (700,300) {Hawking}
\put (700,100) {Island rule}
\end{overpic}
\end{center}
\caption{Page curve (assuming a large initial mass $\m  \gg 1$). \label{fig:page}}
\end{figure}
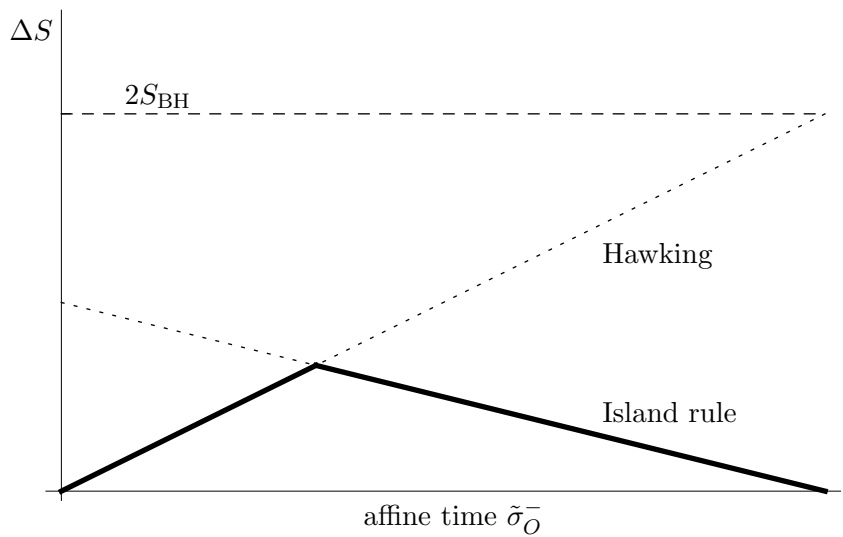

The change in the entropy (which subtracts the IR and UV divergences), plotted in fig.~\ref{fig:page}, has the form anticipated by Page \cite{Page:1993wv,Page:2013dx}: It grows linearly until an O(1) fraction of the black hole has evaporated, then decreases back to zero. 
The Page time $t_P={4\m \over 3}$, when the entropy starts to decrease, is at the point where 1/3rd of the black hole has evaporated. This is because black hole evaporation is non-adiabatic, and in the RST model,  entropy decrease $-\delta S$ of the black hole produces an entropy $+2\delta S$ in the radiation. Therefore the black hole entropy is equal to the radiation entropy after 1/3rd of the black hole evaporates.

\subsection{Scrambling time}
\newcommand{\tsigma}{\tilde{\sigma}}
We now  compute the scrambling time associated to the island rule entropy following \cite{Penington:2019npb, Almheiri:2019psf}. 
The scrambling time is the time it takes for information dropped into a black hole to reappear in the Hawking radiation. The island rule effectively asserts that the information in the island 
at time $x^-_Q$ is available at ${\cal I}^+$ at the canonically associated retarded time $x^-_O$ in \eqref{xqo}. The dropped information hence becomes available when it hits the QES curve. 
 The main difference in asymptotically flat space compared to AdS is that we need to keep track of the initial time in the scrambling experiment. The result agrees with Hayden and Preskill \cite{Hayden:2007cs} (see also \cite{Sekino:2008he,Shenker:2013pqa}), namely $t_{\rm scr} \sim \frac{\beta}{2\pi}\log \frac{S}{N}\sim \log ({\m})$ for large $N$ and $\m$.

\begin{figure}
\begin{center}
\includegraphics[scale=1.0]{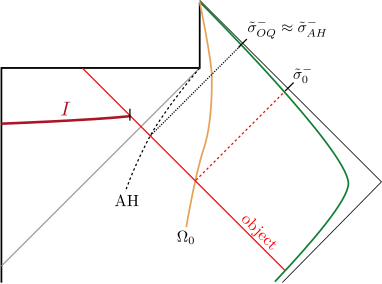}
\end{center}
\caption{\small Scrambling experiment. \label{fig:scrambling}}
\end{figure}

The experiment is illustrated in fig.~\ref{fig:scrambling}. After the Page time, an object is dropped into the black hole by a distant observer, who is at rest in the inertial coordinates near infinity,
\be
\tilde{\sigma}^+ = t + y , \quad \tilde{\sigma}^- = t - y \ .
\ee
The observer should be far from the black hole, but close enough so the black hole is effectively stationary for the whole experiment.
The observer collects the Hawking radiation along ${\cal I}^+$. Let $\tsigma^-_0$ be the retarded time when the object reaches a few Schwarzschild radii from the black hole (the precise definition is given below) and $\tsigma_{OQ}^-$ be the retarded time associated via 
\eqref{xqo} to the object crossing the QES curve and entering  the  island. Then the scrambling time is defined by\footnote{Note that light rays emitted by the object when it crosses the apparent horizon hover near the black hole for a long time and suffer a great delay before arriving at infinity. Hence we cannot equate the scrambling time with the difference between this retarded crossing time $\tsigma_{AH}^-$ and $\tsigma_{OQ}^-$. In fact that difference equals $-\log 3$ and is negative! See \cite{Penington:2019npb,Almheiri:2019psf,Almheiri:2019hni,Almheiri:2019yqk} for discussion.}
\be\label{tmdef}
t_{\rm scr} = \tilde{\sigma}^-_{OQ} - \tsigma^-_0 \ .
\ee
To see that this definition makes sense, note that the experiment has three stages: First the object falls to near the black hole in time $t_{\rm fall}$, then it is scrambled, then the Hawking radiation carrying information about the object takes the same time $t_{\rm fall}$ to return from the near region to the observer. Thus the total proper time elapsed at infinity, from dropping the object to seeing it return in the Hawking radiation, should be 
\be
\Delta t = t_{\rm scr} + 2 t_{\rm fall} \ .
\ee
Since the black hole is effectively stationary, we can see that this definition of the scrambling time implies \eqref{tmdef} upon using $\tsigma^-_0 = t_{\rm obj}+2t_{\rm fall} - y$. 

Now we need to define `near' the black hole.   In higher dimensions this would be a region within a few Schwarzschild radii. In two dimensions, the coordinate-invariant analogue is to define the near region as
\be
\Omega \lesssim \Omega_0 \equiv (1+A) \Omega_H \ ,
\ee
where $\Omega_H$ is $\Omega$ at the horizon, and $A > 0$ is an $O(1)$ constant. We want $A$ to be large so that we can neglect black-hole-related redshifts, but to be parametrically small compared to $N$ or $\m$. Thus we define $\tsigma^-_0$ to be the retarded time when the object reaches the curve $\Omega = \Omega_0$. This is when the observer along ${\cal I}^+$ starts the timer to define the scrambling time.

We take the object to fall on a null trajectory with $x^+ = x^+_{\rm obj}$. We can treat the black hole as stationary, with
\be
\Omega \approx -  x^+ (x^-+M) + \frac{6S_{BH}}{N} \ .
\ee
where as we have seen $\frac{6S_{BH}}{N}=\m$.  The horizon is at $x^-=-M$, i.e. $\Omega = \frac{6S_{BH}}{N}$, so we have $\Omega_0 = (1+A)\frac{6S_{BH}}{N}$.   The object reaches the curve $\Omega = \Omega_0$  at 
\be
\tsigma^-_0 =  \log \left( \frac{N  x_{\rm obj}^+}{6A S_{BH}} \right) \ .
\ee
According to \eqref{approxQ}, the object is seen at ${\cal I}^+$ to enter the island at the retarded time 
\be
 { \tsigma^-_{OQ}}=\log \frac{4x^+_{\rm obj}}{3}.
\ee
Therefore 
\be
\tsigma^-_{OQ} = \tsigma_0^- + \log\left( \frac{8AS_{BH}}{N} \right) \ .
\ee
The $\log(8A)$ term is subleading, so for the scrambling time we find to leading order 
\be
t_{\rm scr} = \frac{\beta}{2\pi}\log \frac{S_{BH}}{N} \ ,
\ee
where the inverse temperature of the black hole is $\beta ={2\pi}{}$. This agrees with \cite{Hayden:2007cs,Sekino:2008he}, including the $\log N$ correction also found in \cite{Penington:2019npb}.

\section{The eternal black hole}
\begin{figure}
\begin{center}
\includegraphics[scale=1.0]{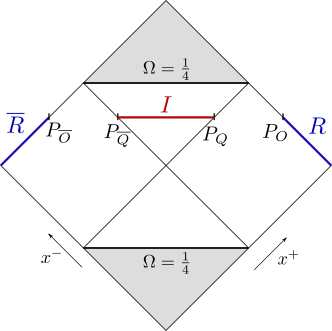}
\end{center}
\caption{\small Island in the eternal black hole. With the radiation collected on ${\cal I}^+$, the endpoints of the island are on the horizon. \label{fig:eternalBH}}
\end{figure}

In this section we apply the island rule to the eternal black hole in the RST model. In addition to being mathematically simpler than the evaporating black hole, the late time behavior of the quantum extremal surfaces in the evaporating black hole is well-approximated by those in the eternal black hole. 
Moreover, as discussed in \cite{Fiola:1994ir,Strominger:1993yf,Almheiri:2019yqk}, the eternal black hole furnishes a simple version of the information paradox: At late times, the black hole in the Hartle-Hawking vacuum seems to have an arbitrarily large entanglement with the Hawking radiation, despite the fact that it has a fixed, finite Bekenstein-Hawking entropy.  Our discussion here follows a closely related AdS version in \cite{Almheiri:2019yqk}.

Consider the entropy in the region $R \cup \overline{R}$, where $R$ is near ${\cal I}^+$ and $\overline{R}$ is its reflection on the other side of the Penrose diagram. (See fig.~\ref{fig:eternalBH}.) Both intervals are of equal size, and as we make them bigger the entanglement entropy increases without bound. This can be seen as follows. Since the global quantum state is pure we consider the entropy on the complement, which is a single interval extending from the point $\overline{P}_O$ on the left to $P_O$ on the right, with both points near ${\cal I}^+$.  The entanglement entropy of a conformal field theory on an interval $[\bx_O, x_O]$ in the spacetime $ds^2 = -e^{2\rho_K}dx^+ dx^-$ is 
\be\label{srr1}
S_{\rm ent}({R\cup \overline{R}}) = \f N 6 \log \left[\f{(x_O^+-\bx_O^+)(\bx_O^--x_O^-)}{\euv ^2e^{-\rho_K(x_O)-\rho_K(\bx_O)}}\right].
\ee
 In the rest of this section, we set $\sigma_O^{\pm} = t_O \pm y_O$, expand in the regime $M \gg 1, t_O \gg 1, y_O \gg \log M, {1 \over \euv}\gg 1$, and drop terms of $O(1)$.  Considering the interval of interest in the black hole background, the entanglement entropy becomes \be\label{noisland}
S_{\rm ent}({R\cup \overline{R}}) \approx  
\frac{N}{3}t_O  - \frac{N}{3}\log \epsilon_{\rm uv}\,.
\ee
This increases linearly with time. On the other hand, unitarity suggests\footnote{The entanglement entropy  \eqref{noisland} could only be measured by metaobservers who access both asymptotic regions of the eternal geometry. Hence the paradox here is less physical than, though a close mathematical analogy to, the one encountered in the previous section or discussed in  \cite{Fiola:1994ir,Strominger:1993yf}.} that this be  less than  the entropy of the black hole together with the thermal bath on $B \cup \overline{B}$, where $B$ is the region between the black hole and $R$: 
\be\label{ubound1}
S(R \cup \overline{R}) \lesssim 2\left( S_{\rm grav} + S_{\rm ent}(B)  \right) \ ,
\ee
with the gravitational term evaluated at the horizon. This is the entropy of the black hole plus bath, with a factor of 2  to account for the two sides. The right-hand side of \eqref{ubound1} is time-independent, so we can calculate it at $t=0$, taking $B$ to be the region from the bifurcation point to $y_O$. Then
\begin{align}
S_{\rm grav} + S_{\rm ent}(B) &\approx
 \Omega_H + \frac{N}{6}\ln \left[ \frac{-x_O^+ x_O^-}{\epsilon_{\rm uv}e^{-\rho_K(P_O)}}\right]\\
&\approx \frac{NM}{6}  + \frac{N y_O}{6} - \frac{N}{6}\log \epsilon_{\rm uv} \ ,
\end{align}
up to subleading terms not proportional to $M$ or $y_O$. Thus the expected unitarity bound \eqref{ubound1} is
\be\label{eternalb}
S(R \cup \overline{R}) \lesssim \frac{NM}{3} + \frac{Ny_O}{3} - \frac{N}{3} \ln \epsilon_{\rm uv} \ .
\ee
There may be subleading corrections to this bound, but in any case eventually it will be overtaken by the linear growth in \eqref{noisland}.  Hence the explicit expression \eqref{noisland} clearly violates \eqref{ubound1}.

Now we consider the island rule entropy. We are looking for a quantum extremal surface which is two points bounding an island containing the interior of the black hole. Then the matter entanglement is computed on the surface $\overline{R} \cup I\cup R$ where $I$ is the island. By purity of the global state we can again consider the entropy on the complement region, which is two intervals $B \cup \overline{B}$. At late times these two intervals are far apart and the total  entanglement entropy becomes twice the single interval answer.  We then have
\be
S_{\text{ent}}({\overline{R}\cup I \cup R}) = \f{N}{3}\log \left[\f{(x_O^+-x_Q^+)(x_Q^--x_O^-)}{\euv ^2 e^{-\rho_K(P_O)-\rho_K(P_Q)}}\right].
\ee
The gravitational entropy also has two contributions, so is given by twice the formula \eqref{sgrav} evaluated at the point $x_Q$.
The generalized entropy is 
\begin{align}
S_{\text{gen}}(\overline{R}\cup I\cup R) &= S_{\text{grav}} + S_{\text{ent}}   \\
&=\f N 3 \left(\Omega(P_Q)+\log \left[\f{(x_O^+-x_Q^+)(x_Q^--x_O^-)}{\epsilon_{\rm uv}}\right] + \rho_K(P_O) + \log 2 - \f 1 4\right)\, . \notag
\end{align}
Extremizing with respect to $x_Q^\pm$ gives 
$x_Q^\pm \approx -\f{1}{ x_O^\mp}$ at large $x_O^+$. In this approximation we find that the endpoint of the island lies on the horizon at
\be
\sigma_Q^+ = \sigma_O^- \ .
\ee
The entropy becomes 
\begin{align}\label{seternali}
S_{\text{gen}} \approx 2S_{\rm BH} + \frac{Ny_O}{3} - \frac{N}{3} \ln \epsilon_{\rm uv} \ .
\end{align}
Comparing to the case with no island \eqref{noisland} shows that the island saddle dominates for 
\be
t_O \gtrsim  {\m}{}  +y_O
\ee
After the Page transition, the entropy saturates the unitarity bound \eqref{eternalb} up to subextensive terms.

\ \ 

\ \ 

\ \ 

\noindent \textbf{Acknowledgments}

\noindent We thank Ahmed Almheiri, Raphael Bousso, Kanato Goto, Dan Harlow, Dan Jafferis, Raghu Mahajan, Juan Maldacena, Suvrat Raju, Douglas Stanford and Amir Tajdini for useful conversations. The work of ES is supported by Simons Foundation grant 488643, the Simons Collaboration on the Nonperturbative Bootstrap. The work of TH is supported by DOE grant  DE-SC0014123. The work of AS is supported by DOE  grant DE-SC/0007870 and by the Gordon and Betty and Moore Foundation and John Templeton Foundation grants via the Black Hole Initiative.

\appendix

\section{Local analysis of replica wormholes}

For black holes in AdS$_2$, the island rule has been derived directly from the gravitational path integral \cite{Almheiri:2019qdq,Penington:2019kki}. The path integral is used to calculate the replica partition functions
\be
Z_n = \tr (\rho_A)^n \ .
\ee
This can be analytically continued to non-integer $n$ by continuing the solution itself, and used to obtain the von Neumann entropy,
\be\label{replicaS}
S(\rho_A) =  - \p_n \left( \frac{Z_n}{(Z_1)^n} \right)_{n=1} \ .
\ee
It is not obvious whether these results carry over to asymptotically flat spacetime, where gravity is dynamical everywhere. In this appendix we make a step in this direction. We discuss replica wormhole geometries for the eternal black hole in the RST model, construct the solutions locally near the quantum extremal surface, and show that the equations of motion enforce the QES condition. However we will not construct the wormhole saddles globally, so this is not a complete derivation of the island rule.

The local analysis closely follows the derivation of the Ryu-Takayanagi formula by Lewkoywycz and Maldacena \cite{Lewkowycz:2013nqa} and the general analysis of quantum extremal surfaces by Dong and Lewkowycz \cite{Dong:2017xht} (see also \cite{Dong:2016hjy}). By assuming a large-$N$ matter sector, we sidestep some of the difficulties in defining the matter entanglement entropy in \cite{Dong:2017xht}.

\subsection{Replica geometries}

We will focus on the eternal black hole at $t=0$. The same results apply (locally near the QES) to any spacetime with a time-reflection symmetry, such that it can be prepared by a Euclidean path integral.

Take $R$ to be the semi-infinite region $y \in [y_O, \infty)$, with $y_O \gg 1$, on the right side of the Penrose diagram for the eternal black hole. This is analogous to the region where an observer collects the Hawking radiation, but it is at $t=0$. Its reflection across the Penrose diagram is $\bar{R}$. We are interested in the von Neumann entropy of $R \cup \bar{R}$, and the associated replica partition functions
\be
Z_n = \tr (\rho_{R \cup \bar{R}})^n \ ,
\ee
calculated from the gravitational path integral.

Consider first $n=1$. The partition function $Z_1 = \tr \rho_{R \cup \bar{R}}$ is independent of the region. The path integral has asymptotic boundary conditions specifying that we have an $S^1$ of size $2\pi$, and the dilaton asymptotics given by \eqref{eternalomega}.\footnote{The canonical ensemble is ill-defined in the RST model, since the temperature is not a free parameter. We therefore fix both the leading and subleading terms in $\Omega$. As usual, boundary terms and counterterms are needed to make the action finite, but we will not need the details.} The saddlepoint is simply the Euclidean black hole, \textit{i.e.}, the cigar geometry:
\begin{align*}
 \vcenter{\hbox{\includegraphics{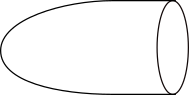}}}
\end{align*}
Next we turn to $n=2$. The standard replica manifold has the same topology as two copies of the Euclidean black hole, glued along the regions $R$ and $\bar{R}$:
\begin{align*}
\vcenter{\hbox{\includegraphics{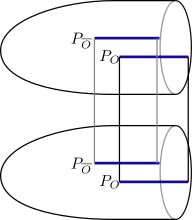}}}
\end{align*}
Each cigar is associated to a factor of $\rho_{R\cup \bar{R}}$, and the gluing accounts for the matrix multiplication in $\tr (\rho_{R \cup \bar{R}})^2$. The gluing is defined cyclically, so the top of region $R$ in copy 1 is glued to the bottom of region $R$ in copy 2, and vice-versa. Note that $\bar{R}$ is half way around the thermal cycle from $R$. There are two distinct asymptotic $S^1$ boundaries. To find the saddle, the equations of motion should be solved everywhere except at the branch points $P_O$ and $P_{\bar{O}}$, where there are $2\pi$ conical excesses.

Still at $n=2$, we can also consider higher topology,  contributions to the path integral, connected in the interior of the geometry:
\begin{align*}
\includegraphics{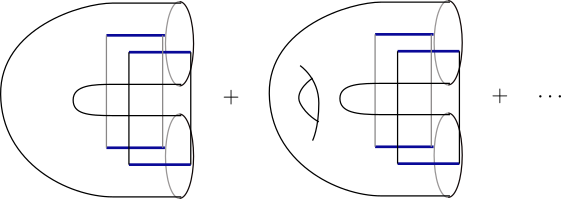}
\end{align*}
 We impose the same boundary conditions as before ---  two copies of $S^1$ and the appropriate dilaton asymptotics, with $2\pi$ conical excesses at $P_O$ and $P_{\bar{O}}$.\footnote{Away from $n=1$ there is an ambiguity in where we place the point $P_O$, because it is in a different metric. This ambiguity would need to be fixed to fully specify the replica saddles, but it will not affect our discussion near $n \sim 1$ as it only appears at $O((n-1)^2)$.} The interior of the manifold, including the wormhole connecting the two replicas, is completely smooth by the equations of motion.

Proceeding to higher $n$, we can in principle consider all topologies at each $n$. The corresponding saddles, if they exist, are the replica wormholes. We will restrict to manifolds with a $\mathbf{Z}_n$ cyclic symmetry permuting the $n$ copies of the cigar. The geometry of a $\mathbf{Z}_n$-symmetric replica wormhole can be represented as a branched cover of a single cigar, with an even number of $n$-fold twist defects:
\begin{align*}
\includegraphics{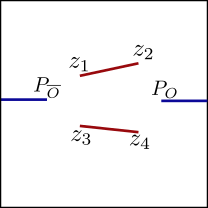}
\end{align*}
Here we have switched to a planar coordinate, so the metric is asymptotically $dz d\bar{z}/(z\bar{z})$, the thermal cycle is $z \to z e^{2\pi i}$,  and the horizon (for $n=1$) is at $z=0$. The full manifold ${\cal M}_n$ has $n$ copies of this picture, glued cyclically along all of the cuts. This full manifold is smooth everywhere except at the \textit{original} branch points $P_O$ and $P_{\bar{O}}$, where it has $2\pi (n-1)$ conical excesses. In particular it is smooth at the new, dynamical branch points $z_i$. Here the equations of motion impose
\be\label{defectzi}
ds^2 \approx \frac{1}{n^2}|z-z_i|^{2/n-2} dz d\bar{z} \quad \mbox{as}\quad z \to z_i \ .
\ee
That is, $ds^2 \approx dw d\bar{w}$ in the coordinate $w = (z-z_i)^{1/n}$ that covers a neighborhood of the defect, so the full manifold is smooth at this point, as required.

The Euclidean constraint equations of the RST model in conformal gauge $ds^2 = e^{2\rho}dz d\bar{z}$ are
\be\label{fullcon}
\left( e^{-2\phi} + \frac{N}{48} \right) (4 \p \rho \p \phi - 2 \p^2 \phi) -\pi T_{zz} = 0 \ ,
\ee
and similarly with bars. Here $T_{zz}$ is the full stress tensor, including the anomaly. The metric dependence can be extracted by writing
\be
T_{zz} = T_{zz}^{\rm flat}  -  \frac{N}{12 \pi}(\p \rho \p \rho - \p^2 \rho)  \ ,
\ee
where $T_{zz}^{\rm flat}$ is the stress tensor in the metric $dz d\bar{z}$. $T_{zz}^{\rm flat}$ is the matter stress tensor on the replica manifold with metric $dz d\bar{z}$. It is therefore the same stress tensor that would appear in a purely CFT calculation of the Renyi entropy, without gravity. The other equations of motion are
\be\label{nneom}
\p \bar{\p} \Omega = \p \bar{\p} \chi = e^{2(\chi - \Omega)} \ .
\ee
$\rho$ is regular near the original defects $P_O$ and $P_{\bar{O}}$. According to \eqref{defectzi}, near the dynamical defects, it obeys
\be\label{rhonn}
\rho \approx \left( \frac{1}{n} - 1\right)\log |z-z_i| \quad \mbox{as}\quad z \to z_i \ .
\ee
To summarize, a $\mathbf{Z}_n$-symmetric replica wormhole is a solution to the equations \eqref{fullcon} and \eqref{nneom}, subject to \eqref{rhonn} at the defects, with the Euclidean black hole boundary conditions at infinity. The fields are also required to be single-valued on the quotient manifold, \textit{i.e.}, on the $z$-plane. As we will see, both the positions of the defects $z_i$  and the behavior of $\phi$ near these points are fixed by the EOM. 

Note that this formulation makes sense for non-integer $n$. This is the advantage of working on the quotient space.

\newcommand{\bz}{\bar{z}}
\newcommand{\bw}{\bar{w}}
\subsection{Derivation of the QES}

We will not attempt to solve the full replica equations in this paper, but we will solve them locally near the defects, for $n \sim 1$. We will see that this fixes the $z_i$ to sit at extrema of the generalized entropy. 

Let us shift the coordinates so that a dynamical defect sits at $z=0$. We will solve the constraint equation \eqref{fullcon} (and its barred version) near $z\sim 0$ to leading order in $n-1$. Expand
\be
\rho \to \rho + (n-1)\delta \rho , \quad
\phi \to \phi + (n-1)\delta \phi , \quad
T_{\mu\nu}^{\rm flat} \to T_{\mu\nu}^{\rm flat} + (n-1)\delta T_{\mu\nu}^{\rm flat} \ .
\ee
The boundary condition \eqref{rhonn} near the defect is
\be\label{rhobc}
\delta \rho \sim -\frac{1}{2}\log (z\bz ) \ .
\ee
The stress tensor $T^{\rm flat}_{\mu\nu}$ is the same stress tensor that appears in the CFT calculation of entanglement entropy in flat space, without gravity \cite{Holzhey:1994we,Calabrese:2004eu}. It satisfies a conformal Ward identity that relates its singular behavior to the CFT entanglement entropy \cite{Calabrese:2004eu},
\begin{align}
2\pi \delta T_{zz}^{\rm flat} &= \frac{N/12}{z^2} - \frac{\p S_{\rm CFT}^{\rm flat}}{z} + \mbox{reg.} \\
2\pi \delta T_{\bz \bz}^{\rm flat} &= \frac{N/12}{\bz^2} - \frac{ \bar{\p} S_{\rm CFT}^{\rm flat}}{\bz} + \mbox{reg.}
\end{align}
In the $1/z^2$ term we used the chiral scaling dimension of the twist operator,
\be
h_n = \bar{h}_n = \frac{N}{24}(n-1/n) \ .
\ee
$S_{\rm CFT}^{\rm flat}$ is the matter entanglement entropy in the metric $dz d\bz$.  Note that it depends implicitly on the global structure of the replica manifold. In particular it depends on the locations of $P_O$, $P_{\bar{O}}$, and all of the  dynamical defects $z_i$.

Now we will show that the constraints near $z \to 0$ fix the position of the defect. The dilaton has an expansion
\begin{align}
\delta \phi &= a_{00} + a_{10}z + a_{01} \bz + a_{11}z \bz + a_{20} z^2 + a_{02} \bz^2 +  \cdots\\
&\quad \log(z\bz)(b_{00} + b_{10}z + b_{01} \bz + b_{11}z \bz + b_{20} z^2 + b_{02} \bz^2 +  \cdots) \notag
\end{align}
Note that terms like $z \log z$ are not allowed, except in the combination $z \log (z \bz)$. This would not be single valued on the replica manifold. Plug this expansion into \eqref{fullcon} and expand $z,\bz \to 0$. The leading quadratic singularity fixes $b_{00}$. The simple pole term in the $zz$ constraint fixes $b_{01}$ and $b_{10}$. The we plug the results into the $\bz \bz$ constraint and from the simple pole find the conditions
\be\label{gotqes}
\p( S_{\rm BH} + S_{\rm CFT} ) = \bar{\p} (S_{\rm BH} + S_{\rm CFT}) = 0 \ .
\ee
Here
\be
S_{\rm BH} = 2e^{-2\phi} - \frac{N}{12} \phi + \mbox{const}. 
\ee
is the gravitational contribution to the entropy \cite{Fiola:1994ir,Myers:1994sg}, 
and
\be
S_{\rm CFT} = S_{\rm CFT}^{\rm flat} + \frac{N}{6}\rho(P_{QES}) + \dots \,
\ee
This last equation is the Weyl transformation of the entropy from $dz d\bz \to e^{2\rho} dz d\bz$. The dots refer to the Weyl factors at all other twist points, which do not affect the extremization \eqref{gotqes}. 

Thus we have derived the position of the QES directly from the replica equations of motion. It would be very interesting to find global solutions in Lorentzian signature, with $P_O$ and $P_{\bar{O}}$ taken to ${\cal I}^+$.

\small
\bibliographystyle{ourbst}
 \bibliography{rst-paper}
\end{document}